\documentclass[pra,twocolumn,superscriptaddress,floatfix,amsmath,nofootinbib,amssymb]{revtex4-1}
\usepackage[UKenglish]{babel}
\usepackage{graphicx}
\usepackage{dcolumn}
\usepackage{bm}
\usepackage{verbatim}
\usepackage{float}

\usepackage{mathrsfs}
\usepackage{color}
\newcommand{\ket}[1]{\left| {#1} \right\rangle}
\newcommand{\bra}[1]{\left\langle {#1} \right|}

\newcommand{\proj}[2]{\left| {#1} \right\rangle\!\left\langle {#2} \right|}

\newcommand{\tr}{\operatorname{Tr}}

\newcommand{\eq}[1]{(\ref{#1})}

\newcommand{\qr}{{q_\text{R}}}
\newcommand{\ql}{{q_\text{L}}}

\begin{document}

\title{Convergence of fermionic field entanglement at infinite acceleration\\ in relativistic quantum information}
\author{Miguel Montero}
\affiliation{Instituto de F\'{i}sica Fundamental, CSIC, Serrano 113-B, 28006 Madrid, Spain}
\author{Eduardo Mart\'{i}n-Mart\'{i}nez}
\affiliation{Institute for Quantum Computing, University of Waterloo, 200 Univ. Avenue W, Waterloo, Ontario N2L 3G1, Canada}

\begin{abstract}We provide a simple argument showing that, in the limit of infinite acceleration, the entanglement in a fermionic field bipartite system must be independent of the choice of Unruh modes. This implies that most tensor product structures used previously to compute field entanglement in relativistic quantum information cannot give rise to physical results.\end{abstract}

\maketitle

\section{Introduction}
Recently, \cite{Mig2} showed that previous works in relativistic quantum information \cite{AlsingSchul,Edu2,Edu3,chapucilla,chapucilla2,Geneferm,chor1,Edu9,Edu10,chor2,Mig1} had a flaw in the way they computed entanglement measures for fermionic fields. There has been a debate about what is the proper way to deal with fermions in Relativistic Quantum Information. After the publication of \cite{Mig2}, a comment on this article appeared \cite{BradlerC}, and the arguments were contested in \cite{Bradans}. 

The present work constitutes both a simple argument supporting the claims of \cite{Mig2} and also an interesting observation about entanglement measures for fermionic fields in noninertial frames. Without assuming any mapping between fermions and qubits we show that entanglement measures must behave in a particular way in the infinite acceleration limit.  Any technique to calculate entanglement measures in fermionic fields in the frame of relativistic quantum information is bound to reflect this behaviour.

In this brief report, we provide a general argument showing that in the limit of infinite acceleration, the remaining entanglement must be independent of the choice of Unruh modes. This is not the case in previous works, where field entanglement at infinite acceleration was found to be dependent on the choice of Unruh modes. This is because in previous works did not take into account the correct tensor product structure, as explained in \cite{Mig1} and \cite{Bradans}. Convergence of fermionic entanglement at infinite acceleration is a necessary condition (for the special family of states considered in the literature) for obtaining correct, physical results.

Indeed, the correct result in the asymptotic limit is well recovered if the so-called `physical ordering' defined in \cite{Mig2} is used to endow the fermionic space with a tensor product structure, giving a simple argument in favour  of the correctness and feasibility of the physical mapping implemented in \cite{Mig2}. 

This brief report is structured as follows: In section \ref{set} we briefly introduce the setting and notation. Section \ref{proof} contains our main statement and the corresponding proof. Finally, section \ref{cojoniak} contains ours conclusions.

\section{Setting}\label{set}
\begin{figure}[hbtp] 
\includegraphics[width=.50\textwidth]{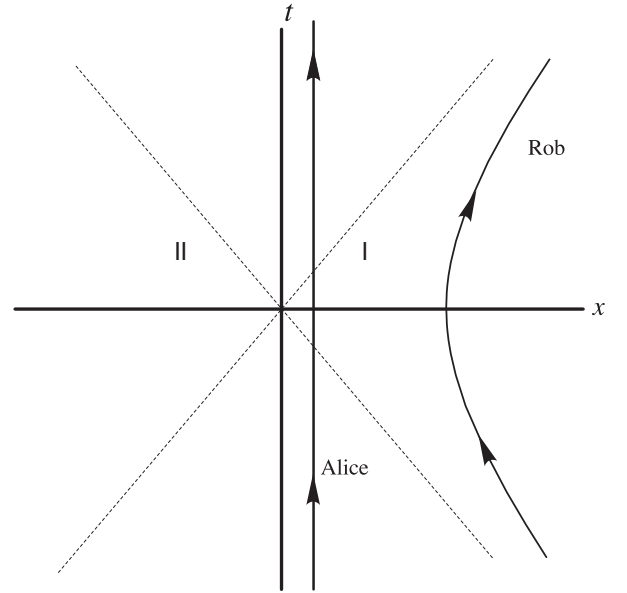}
\caption{Minkowski spacetime diagram showing the world lines of an inertial observer Alice, and one uniformly accelerated observer moving hyperbolically in region I. Note that regions I and II are causally disconnected from each other.}
\label{rindler}
\end{figure}

We consider a fermionic field of spin $s$ and the same setting as in \cite{Mig2} (illustrated in Fig.\ref{rindler}), where a  bipartite field state was shared between an inertial observer, Alice, and a uniformly accelerated one, Rob. Rob follows a worldline of fixed spacelike Rindler coordinate, and therefore uses Rindler modes to describe his part of the field state. 

The field states considered will be the same as in  \cite{Mig2}, namely
\begin{align}\ket{\Psi}&=P\ket{0}_\text{A} (A^\dagger_\text{U}\ket{0}_\text{U})+Q\ket{1}_\text{A} (B^\dagger_\text{U}\ket{0}_\text{U}),\label{est}\\\nonumber  &\quad \vert P\vert^2+\vert Q\vert^2=1,\end{align}
where $A_\text{U}$ and $B_\text{U}$ are arbitrary linear combinations of products of Unruh modes $C^\dagger_{\sigma,\text{U}}$, where
\begin{align}C^\dagger_{\sigma,\text{U}}&=\qr C^\dagger_{\sigma,\text{R}}+\ql C^\dagger_{\sigma,\text{L}},\quad \vert\qr\vert^2+\vert\ql\vert^2=1,\nonumber\\
C^\dagger_{\sigma,\text{R}}&=\cos r\, c^\dagger_{\sigma,\text{I}}- \sin r\, d_{-\sigma,\text{II}},\nonumber\\
C^\dagger_{\sigma,\text{L}}&=\cos r\, c^\dagger_{\sigma,\text{II}}- \sin r\, d_{-\sigma,\text{I}}.\label{umodes}\end{align}
$r$ is a parameter that accounts for the Unruh mode studied and the acceleration of the non-inertial observer \cite{Edu9}. For our purposes, it is enough to say that in the limit of infinite acceleration $r\rightarrow \pi/4$ for all possible Unruh modes.  $c^\dagger_{\sigma,\text{I(II)}}$ are Rindler particle creation operators for spin $z$-component $\sigma$ and spacetime region I (or II). The vacuum $\ket{0}_\text{U}$ is the Unruh vacuum, annihilated by all the Unruh modes. Since these modes are purely of positive frequency in terms of Minkowski modes, it follows that the Minkowski and Unruh vacuums coincide. Similar conventions apply to the antiparticle modes, $d^\dagger_{\sigma,\text{I(II)}}$.

To study entanglement between Alice's and Rob's field modes, first one has to trace over the Rindler modes causally disconnected from the accelerated observer Rob. If Rob's worldline lies in region I, as depicted in Fig. \ref{rindler}, we will have to trace over region II modes, and vice-versa. After this is done, entanglement of the reduced state may be studied for instance by computing entanglement measures such as negativity. This is the standard procedure used for studying bipartite field entanglement in relativistic quantum information \cite{Alicefalls,AlsingSchul}

\section{The main result}\label{proof}
We shall give a full proof of our result only for the Grassman scalar field, an anticommuting field with only one degree of freedom. This means that the $\sigma$ label indexing spin may be dropped. The proof may be extended straightforwardly to higher spin fields. 

In the proof below we do not carry out any mapping from fermionic states to qubits. This general proof shows that any physical procedure that evaluates negativity should be independent of the choice of Unruh modes. 

The equation \eq{est} for the Grassman case reads
\begin{align}\ket{\Psi}&=P\ket{0}_\text{A}\left[a_1\mathbf{I}+a_2C^\dagger_\text{U}\right]+Q\ket{1}_\text{A}\left[b_1\mathbf{I}+b_2C^\dagger_\text{U}\right] \ket{0}_\text{U}\label{est2},\\\nonumber &\vert P\vert^2+ \vert Q\vert^2= \vert a_1\vert^2+ \vert a_2\vert^2= \vert b_1\vert^2+ \vert b_2\vert^2=1.\end{align}

At $r=\pi/4$ (infinite acceleration limit) the Unruh mode \eq{umodes} can be written as
\begin{align}c_\text{U}^\dagger=\left[\qr c^\dagger_\text{I}-\ql d_\text{I}\right]+\left[\ql c^\dagger_\text{II}-\qr d_\text{II}\right]=a^\dagger_\text{I}+a^\dagger_\text{II}\end{align}
where we have defined the modes
\begin{align}a^\dagger_\text{I}&=\qr c^\dagger_\text{I}-\ql d_\text{I},\nonumber\\ a^\dagger_\text{II}&= \ql c^\dagger_\text{II}-\qr d_\text{II}.\label{amodes}\end{align}
The expression for the Unruh vacuum in terms of the Rindler vacuum can be found elsewhere \cite{Edu9} and it is
\begin{align}\ket{0}_\text{U}&=\frac12\left(\cos^2r\,\mathbf{I}+\cos r \sin r\,c^\dagger_{\text{II}}d^\dagger_{\text{I}}-\cos r\, \sin r\,d^\dagger_{\text{II}}c^\dagger_{\text{I}}\right.\nonumber\\&+\left.\sin ^2 r\, d^\dagger_{\text{II}}c^\dagger_{\text{II}}c^\dagger_{\text{I}}d^\dagger_{\text{I}}\right)\ket{0}_\text{Rindler}.\label{uvac}\end{align}

From eqs. \eq{amodes} and \eq{uvac} it is straightforward to check that  at $r=\pi/4$, we have $a^\dagger_\text{I}\ket{0}_\text{U}=a^\dagger_\text{II}\ket{0}_\text{U}$. This means that whenever a region II operator $a^\dagger_\text{II}$ appears in \eq{est2}, we can substitute it by the region I operator $a^\dagger_\text{I}$, since all operators act directly on the Unruh vacuum. Then, any field state being a superposition of the vacuum and Unruh modes such as \eq{est2} can be written as 
\begin{align}\ket{\Psi}=A_\text{I}\ket{0}_\text{U}\end{align}
where $A_\text{I}$ is a linear operator containing only Alice and region I modes. The density matrix of the state is
\begin{align}\proj{\Psi}{\Psi}=A_\text{I}\ket{0}_\text{U}\bra{0}_\text{U}A^\dagger_\text{I}.\end{align}
Now, it is obvious from purely physical considerations that the operator $A_\text{I}$ and its adjoint commute with the tracing over region II modes.  Indeed, these operators do not change the population of region II modes, and therefore, tracing over them can be done before or after applying the operator $A_\text{I}$. But then the relevant reduced state is
\begin{align}\rho=\tr_\text{II}(\proj{\Psi}{\Psi})=A_\text{I}\tr_\text{II}(\ket{0}_\text{U}\bra{0}_\text{U})A^\dagger_\text{I}\propto A_\text{I} A_\text{I}^\dagger\end{align}
where the last equality holds because at $r=\pi/4$ the reduced state of the vacuum is a multiple of the identity (it is a thermal state at infinite temperature). We have shown that when we express the reduced state in terms of the $a_\text{I}^\dagger$ mode instead of the usual Rindler modes, the field state can be expressed without making any explicit reference to $\qr$. 

As the change of basis from the usual Rindler modes to a basis containing $a_\text{I}^\dagger$ is a local unitary operation which does not change entanglement, we have shown that at infinite acceleration the entanglement properties of the state are independent of $\qr$, i.e. of the choice of Unruh mode.

The previous argument made use of the properties of the partial trace operation which stem from its definition as the only operator containing only region I modes  with the same matrix elements between region I observables as the original state would. It is thus independent of the way the reduced state is computed, wether by using fermion-qubit mappings as it was done in previous works \cite{AlsingSchul,Edu2,Edu3,chapucilla,chapucilla2,Shahpoor,Geneferm,chor1,Edu9,Edu10,chor2,Mig1}, or by any other means. 

\begin{figure}[hbtp] 
\includegraphics[width=.46\textwidth]{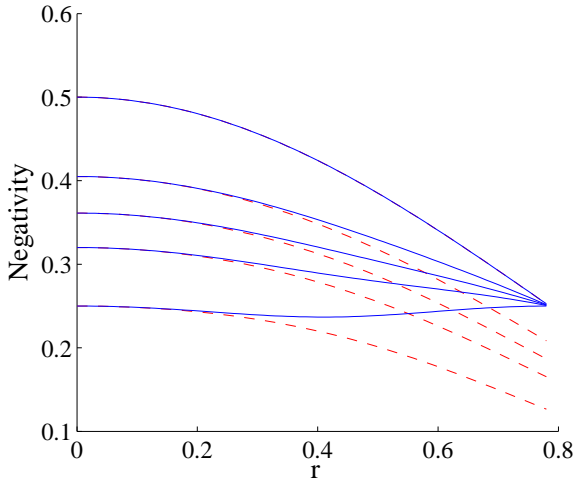}
\caption{Color online: Negativity as a function of $r$ for the state \eq{est2} with $P=Q=1/\sqrt{2}$, $a_1=b_2=1$, $a_2=b_1=0$. Blue solid line corresponds to the physical ordering of \cite{Mig2}, whereas the red-dashed reproduce the results of \cite{Edu9} where the very same field state was analysed employing the unphysical operator ordering used in  \cite{Edu9,Edu10,Mig1}.}
\label{difer}
\end{figure}

As Figure \ref{difer} clearly shows, entanglement convergence at infinite acceleration does not occur for all fermion-qubit mappings. In particular, it does not hold for the mappings used in \cite{Edu9}. This constitutes an explicit proof by example of the fact that not all fermion-qubit mappings (equivalently, tensor product structures on the fermionic Fock space) give rise to correct, physical results. For a more formal proof of this fact, we refer the reader to \cite{Bradans}.

However, in previous literature in fermionic entanglement in non-inertial frames, it was customary to compute entanglement measures after a certain tensor product structure had been endowed on the fermionic system. That is, instead of considering the full fermionic system and take into account all the anticommutation signs that appear when computing the reduced state, these previous works would choose some operator ordering for defining the Fock basis and afterwards treat the system as a collection of qubits, with no anticommutation properties \cite{Mig2,BradlerC}. If not done carefully, this procedure introduces spurious signs which result in an unphysical behaviour for entanglement measures; in particular, there is no convergence of entanglement in the limit of infinite acceleration, as fig. \ref{difer} shows. If the correct mapping (physical ordering described in \cite{Mig3} is used), the correct behaviour is recovered.
 
The results of this brief report make clear that the procedure used by previous works in fermionic entanglement in noninertial frames has to be revised, but there is still a way to use this kind of fermion-qubit mapping without losing the physical results: even though the mapping of the fermionic Fock space to a qubit system does not respect the canonical anticommutation relations \cite{BradlerC}, it can be shown that it is a well-defined mathematical procedure, and the class of physical operator orderings can be rigorously identified \cite{Bradans}.

\section{Conclusions}\label{cojoniak}
We have given a particularly simple proof of the fact that at the infinite acceleration limit ($r=\pi/4$), the residual entanglement of some fermionic field states is independent of $\qr$, i.e. of the choice of Unruh modes. This means that at infinite acceleration there is no difference in working  beyond the single mode approximation ($\qr=1$) or not, at least for this setting.

The relevance of this result lies in the fact that it constitutes a simple requirement that all results in field entanglement in non-inertial frames should fulfill. This requirement was not fulfilled in previous works \cite{Edu9,Edu10,Mig1}, the reason being that all works on field entanglement beyond the single mode approximation so far (except for \cite{Mig2,Mig3}) choose an unphysical operator ordering to study entanglement.

\section{Acknowledgments}

Eduardo Mart\'in-Mart\'inez was supported by a CSIC JAE-PREDOC2007 Grant, the Spanish MICINN Project FIS2008-05705/FIS, and the QUITEMAD consortium.

%

\end{document}